\documentclass[runningheads,envcountsame,envcountsect,oribibl]{llncs}

\usepackage[latin1]{inputenc}
\usepackage{epsfig}
\usepackage{psfrag}
\usepackage{amsmath}
\usepackage{amssymb}
\usepackage[english]{babel}
\makeatletter
\long\def\@doifundefined#1#2{\def\reserved@jhh{\relax}%
 \expandafter\ifx\csname#1\endcsname\relax
    \def\reserved@jhh{#2}%
 \fi
 \reserved@jhh}
\newcommand{\provideenvironment}[2]
  {\@doifundefined{#1}{\newenvironment{#1}#2}}
\newcommand{\providetheorem}[2]
  {\@doifundefined{#1}{\newtheorem{#1}#2}}
\makeatother

\def\rcs$#1${#1}
\newcommand{\version}[1]{\thanks{\rcs#1}}

\newcommand{\call}{a}
\newcommand{\term}[1]{\emph{#1}}		% to define/introduce terms
\newcommand{\etal}{\textit{et al.}~}	% et al.
\newcommand{\ie}{{i.e.},\ }		% i.e.
\newcommand{\Prob}[1]{\mathord{\textbf{Pr}}\left[#1\right]} % probability
\newcommand{\CProb}[2]{\Prob{\left.#1\right| #2}}
\newcommand{\assign}{\mathrel{:=}}	% assignment
\newcommand{\keyw}[1]{\ensuremath{\mathbf{#1}}} % keyword used in program listing
\newcommand{\xor}{\oplus}		% bitwise exclusive or
\newcommand{\Z}{\mathbb{Z}}		% the set of integers
\usepackage{float}
\floatstyle{plain}
\newfloat{protocol}{t}{lop}[section]
\floatname{protocol}{Protocol}

\newcommand{\IF}{\keyw{if}~}
\newcommand{\THEN}{\keyw{then}~}
\newcommand{\ELSE}{\keyw{else}~}

\makeatletter
\newcommand{\@Setstar}[1]{\left\{{#1}\right\}}
\newcommand{\@Set}[2]{\@Setstar{{#1},\ldots,{#2}}}
\newcommand{\Set}{\@ifstar{\@Setstar}{\@Set}}	
\makeatother

\newcommand{\eke}{$\varphi$KE}
\newcommand{\concat}{\parallel}

\makeatletter
\newcommand{\xRightarrow}[2][]{\ext@arrow 0359\Rightarrowfill@{#1}{#2}}
\newcommand{\xLeftarrow}[2][]{\ext@arrow 3095\Leftarrowfill@{#1}{#2}}
\makeatother

\newcommand{\sendright}[1]{$\xrightarrow{\quad #1\quad}$}
\newcommand{\sendleft}[1]{$\xleftarrow{\quad #1\quad}$}
\newcommand{\broadcastright}[1]{$\xRightarrow{\quad #1\quad}$}
\newcommand{\broadcastleft}[1]{$\xLeftarrow{\quad #1\quad}$}

\newcommand{\phase}[1]{
  \multicolumn{3}{l}{\raisebox{-2mm}{\fbox{\textit{#1}}}\hrulefill} \\[3mm]
}

\newcommand{\modp}[1]{#1}

\renewcommand{\cap}{\eta}
\newcommand{\sks}{\sigma}
\newcommand{\sps}{t}
\newcommand{\spl}{s}

\newlength{\protindent}
\title{The Ephemeral Pairing Problem\version{$Id: pairing.tex,v 1.11 2003/11/24 11:34:49 hoepman Exp $}}

\author{Jaap-Henk Hoepman} 
\institute{Department of Computer Science, University of Nijmegen\\
  P.O.Box 9010, 6500 GL \ Nijmegen,
  the Netherlands\\ \email{jhh@cs.kun.nl}}

\begin{document}

\maketitle

\bibliographystyle{alphacm-}

\begin{abstract}
In wireless ad-hoc broadcast networks the \emph{pairing problem} consists of
establishing a (long-term) connection
between two specific physical nodes in the network that do not yet know each
other.  We focus on the \emph{ephemeral} version of this problem. Ephemeral
pairings occur, for example, when electronic business cards are exchanged
between two people that meet, or when one pays at a check-out using a wireless
wallet.

This problem can, in more abstract terms, be phrased as an \emph{ephemeral key
exchange} problem: given a low bandwidth authentic (or private) communication
channel between two nodes, and a high bandwidth broadcast channel, can we
establish a high-entropy shared secret session key between the two nodes
without relying on any a priori shared secret information.

Apart from introducing this new problem, we present several ephemeral key
exchange protocols, both for the case of authentic channels as well as for the
case of private channels.

\end{abstract}

\section{Introduction}

In wireless ad-hoc broadcast networks like 
Bluetooth\footnote{
  See \texttt{http://www.bluetooth.com}.
}
or
IrDA\footnote{
  See \texttt{http://www.irda.org}.
}
there is no guarantee that two physical nodes that want to
communicate with each other are actually talking to each other.  
The \emph{pairing problem} consists of securely establishing
a connection or relationship between two specific nodes in the network that do
not yet know each other\footnote{
  Note the subtle difference with authentication: in the pairing problem we are
  not interested in the actual identity of any of the nodes. In fact, in a
  wired network the problem is easily solved by checking that a single wire
  connects both nodes.
}.
For example, to insure that a newly bought television
set is only controllable by \emph{your} old remote control, the two need to be
paired first. Because this pairing is performed only once (or a few times)
during the lifetime of any pair of nodes, the pairing procedure can be quite
involved.
% (and actually is in many cases).
The importance of pairing, and the security policies governing such long-term
paired nodes, is described by Stajano and Anderson~\cite{StaA99}.

Sometimes, pairings may have to be performed much more frequently, and should
only establish a relationship for the duration of the connection between the
two nodes.  Such \emph{ephemeral pairings} occur, for example, when exchanging
electronic business cards between two people that happen to meet, or when
paying at a check-out using a wireless wallet on your mobile phone. Because
such pairings may happen many times a day, the pairing procedure should be fast
and the amount of user intervention should be limited. On the other hand,
a high level of trust in the pairing may be required. Therefore, 
the pairing should be established in such a way that a high level of security
is achieved even with minimal user interaction. Additionally, privacy may be a
concern. Finally, the pairing should be made on the spot, preferably without
any preparations. 

To achieve such pairings, we do not wish to rely on any secret
information shared a priori among the nodes. For the large scale systems
where we expect the ephemeral pairings to play a part, such a secure
initialisation might be costly and carry a huge organisational burden. Instead,
we allow the nodes in the system to exchange \emph{small} amounts of
information reliably and/or privately. Several realistic methods for doing so
are briefly discussed in this paper.

\begin{figure}[t]
\begin{center}
\epsfig{file=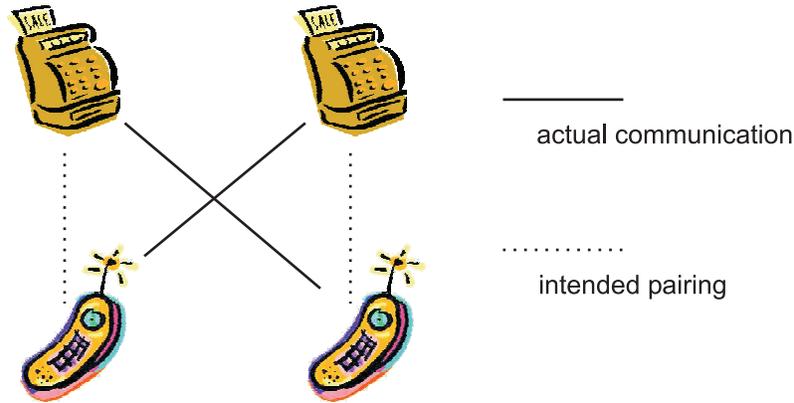}
\end{center}
\caption{Unwanted exchange of information between unpaired nodes.}
\label{fig-swap}
\end{figure}

The importance of correctly pairing nodes becomes apparent if we study the two
examples just given in slightly more detail (see Fig.~\ref{fig-swap}). 
If some people in a crowd start exchanging
business cards that may also contain quite personal information, the business
cards surely should not be mixed up by the wireless network. Similarly, if two
people are about to pay using a wireless wallet at two adjacent check-outs in a
supermarket, the system should make sure that both are paying the right bills.
In fact, similar problems plague smart card purse based systems like
the Common Electronic Purse Specifications (CEPS~\cite{CEPS01}),
see~\cite{JurW01} for details.

The ephemeral pairing problem can also be phrased in more abstract terms as a
key exchange problem. Suppose we are given a low bandwidth authentic (or
private) communication channel between two nodes, and a high bandwidth
broadcast channel, can we establish a high-entropy shared secret session key
between the two nodes without relying on any a priori shared secret
information? We call this problem the \emph{ephemeral key exchange} 
(denoted by \eke{})
problem. Here, the low bandwidth channel models the (implicit) authentication
and limited information processing capabilities of the users operating the
nodes.

\subsection{State of the art}

The ephemeral key exchange problem is related to the encrypted key exchange
(EKE) problem introduced by Bellovin and Merritt~\cite{BelM92,BelM93}. There,
two parties sharing a low entropy password are required to securely exchange a
high entropy session key. For \eke{}, the two parties do not share a password,
but instead can use a small capacity authentic and/or private channel. EKE
protocols are not suitable for this setting directly, most certainly not
when only authentic channels are available. However, using private channels and
with some minor additions they can be used to solve the \eke{} problem.
This relationship is explored further in Sect.~\ref{sec-exchange}.

Jablon~\cite{Jab96} thoroughly discusses other solutions to the EKE problem.
This paper also contains a good overview of the
requirements on and a comparison among different EKE protocols.
A more rigorous and formal treatment of the security of EKE protocols was
initiated by Lucks~\cite{Luc97}, and expanded on by several authors
\cite{BoyMP00,BelPR00,Sho99,CanK01,GenL03}. This was followed by several 
more new proposals for EKE protocols secure in this more formal sense, cf. 
\cite{Mac01a,KatOY01}.

\subsection{Contribution and organisation of this paper}

We first introduce and define the ephemeral pairing problem and the ephemeral
key exchange problem, and show how both are related. To the best of our
knowledge, both problems have never before been studied in the literature.
Next, in Sect.~\ref{sec-exchange}, we present ephemeral
key exchange protocols both for the case where the nodes are connected through
authentic channels and when the nodes are connected using private channels.
In Sect.~\ref{sec-appl} we discuss how such authentic and private channels 
could be implemented in practice.
We discuss our results in Sect.~\ref{sec-concl}.

\section{The ephemeral pairing problem}

Consider $n$ physically identifiable nodes communicating over a broadcast
network\footnote{
  In general the wireless network may not be completely connected and may
  change dynamically during the course of the protocol; we can safely ignore
  these cases, because they do not change the essence of the problem.
}, each attended by a human operator. The operators (and/or the nodes they
operate) can exchange \emph{small} amounts of information reliably and/or in
private. 

The ephemeral pairing problem requires
two of these nodes (to be determined by their operators) to establish a shared
secret such that 
\begin{description}
\item[(R1)] both nodes are assured the secret is shared with the correct physical
  node, 
\item[(R2)] no other node learns (part of) the shared secret, and
\item[(R3)] the operators need to perform only simple, intuitive steps.
\end{description}

The shared secret can subsequently be used to set up a secure channel over the
broadcast network between the two nodes. The generalised ephemeral pairing
problem among $m<n$ nodes requires $m$ nodes to establish a shared secret. We
do not study that problem here. A weaker version of the ephemeral pairing
problem requires only one node (the \emph{master}) to be assured that the other
node (the \emph{slave}) actually shares the secret with it. This is called the
\emph{one-sided} ephemeral pairing problem\footnote{ 
  This applies to the case where the slave is unattended by an operator. A typical
  scenario would be paying with a wireless wallet (the master) at a vending 
  machine (the slave). Note that now the slave has no clue (physically) with
  whom it shares the secret. 
}.

\subsection{Using channels to define the problem}

As explained in the introduction, this problem can be seen in more abstract
terms as an ephemeral key exchange (\eke) problem. In this case, 
Alice and Bob share a low bandwidth communication channel over which they 
can exchange at most $\cap$ bits of information per message\footnote{
  We require that the number of messages exchanged over the channel in a single
  protocol run is constant, and small. This, together with the small size
  of $\cap$ formalises requirement (R3) above.
}. 
This channel is either
\begin{description}
\item[authentic,] meaning that Bob is guaranteed that a message he receives
  actually was sent by Alice (but this message may be eavesdropped by others),
  or 
\item[private,] meaning that Alice is guaranteed that the message she sends
  is only received by Bob (but Bob does not know the message comes from Alice).
\end{description}
These guarantees may hold in both directions, or only in one
direction\footnote{ 
  Note that in the case of an unidirectional authentic channel for solving the 
  one-sided \eke{} problem, the channel runs from the slave to the master. See
  Sect.~\ref{sec-appl} for concrete examples.
}.
We note that the low-bandwidth restriction of both the authentic and the
private channel is important in practice. For instance, an authentic channel
could be implemented by a terminal showing some small number on its public
display, that must entered manually on the other terminal.
Sect.~\ref{sec-appl} discusses more examples of such authentic and private
channels.

\begin{figure}[t]
\begin{center}
\epsfig{file=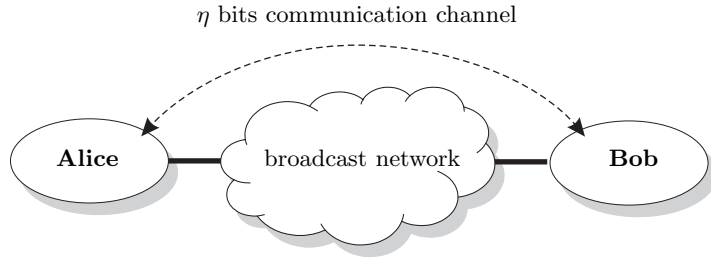}
\end{center}
\caption{The ephemeral key exchange system model.}
\label{fig-exchange}
\end{figure}

Alice and Bob are also connected through a high bandwidth broadcast
network (see Fig.~\ref{fig-exchange}). In this paper, we assume that for the 
correct delivery of broadcast messages, and to separate message streams from
different protocol instances, the nodes on the network have unique identities.
Messages on the broadcast channel carry a small header containing the identity
of both the sender and the receiver. Clearly, the adversary has full control
over the contents of these header fields as well.
\psfrag{alice}{\textbf{Alice}}
\psfrag{bob}{\textbf{Bob}}
\psfrag{ac}{$\cap$ bits communication channel}
\psfrag{bc}{broadcast network}
Given these connections, Alice and Bob are required to establish an
authenticated and shared $\sks$ bits secret (where $\sks \gg \cap$). They do
not share any secrets a priori, and do not have any means to authenticate each
other, except through the low bandwidth channel.

The adversary may eavesdrop, insert and modify packets on the broadcast
network, and may eavesdrop on the authentic channel or insert and modify
packets on the private channel. Note that, by assumption, the adversary cannot
insert or modify packets on the authentic channel.
Also, the adversary may subvert any number of
nodes and collect all the secret information stored there.

\subsection{Model and definitions}

\renewcommand{\inst}[2]{\Pi_{#1}^{#2}}
\renewcommand{\call}[1]{\text{\texttt{#1}}}
\newcommand{\adverse}{{\cal{A}}}
\newcommand{\adv}{\text{\textsf{Adv}}}

We prove security of our protocols in the encrypted key exchange model
developed by Bellare \etal\cite{BelPR00}. For self containment reasons, we
briefly summarise this model here. 

There is a fixed set of principals, that either behave as clients or as
servers. Each principal $p$ may engage in the protocol many times.
Each time
this creates a new, unique, instance $\inst{p}{i}$. Instances of a single
principal share the global state maintained by that principal. This state is
not accessible to the adversary (but see below).

Communication over the network is assumed to be controlled completely by the
adversary. Interaction of the adversary with protocol instances of a principal
is modelled by giving the adversary access to oracles for those instances. 
Let $P$ be the protocol under consideration.
For each instance $\inst{p}{i}$ the following oracles exist.
\begin{description}
\item[$\call{Send}(p,i,m)$]
  Sends or broadcasts message $m$ to instance $\inst{p}{i}$. Any responses or
  output according to $P$ are given to the adversary.
\item[$\call{Execute}(p,i,q,j)$]
  Executes a complete protocol run of $P$ between client $\inst{p}{i}$ and
  server $\inst{q}{j}$. The adversary learns all the messages exchanged between
  the instances, and whether they accept or not.
\item[$\call{Reveal}(p,i)$]
  Reveals the session key generated by instance $\inst{p}{i}$ to the adversary.
\item[$\call{Test}(p,i)$]
  Can be called only once at any time in each execution. A bit $b$ is flipped
  at random, and depending on the outcome the adversary is given either a
  random session key (when $b=0$), or the session key generated by instance
  $\inst{p}{i}$ (when $b=1$).
\end{description}
An execution of the protocol $P$ is defined as a sequence of oracle calls
performed by the adversary. Two instances are called paired\footnote{
  Formally, pairing can be defined as follows. Let the trace of an instance be 
  the concatenation of all messages sent and received by that instance. Then
  two instances are paired when their traces are equal.
} 
if they jointly ran protocol $P$. For a correct protocol, two paired instances
must share the same session key. 

The aim of an adversary $\adverse$ attacking protocol $P$ is to correctly guess
whether the call to the $\call{Test}(p,i)$ query returned the session key
of that instance or just a random session key (or, in other words, to guess the
value of the coin flip $b$ used in the query). Let $S_{\adverse}^P$ denote the
event that adversary $\adverse$ correctly guesses the value of the bit when
attacking protocol $P$. 
Then the advantage of an adversary $\adverse$ attacking protocol $P$ is defined as
follows: 
\[
  \adv_{\adverse}^P = 2 \, \Prob{S_{\adverse}^P} - 1~,
\]
(where $\Prob{X}$ denotes the probability of event $X$).
To make this a non trivial task, the adversary is restricted in the the sense
that it is not allowed to call the $\call{Test}(p,i)$ query if it called the
$\call{Reveal}$ query on $\inst{p}{i}$ or on the instance paired with it.

Each protocol is actually a collection of protocols that must be instantiated
using a particular value for its security parameter. In the case of \eke{}
protocols there are actually two security parameters. There is a 
large security parameter $\spl$ (that roughly corresponds to the size of the
session key to be established, and that mostly determines the advantage of a
passive adversary), and there is a small security parameter $\sps$
(that roughly corresponds to the capacity of the channel between two
principals, and that mostly determines the advantage of an active adversary).

In our analysis we will bound the advantage of the adversary for a particular
protocol using $\spl$, $\sps$ and the number of $\call{Send}$ queries (denoted
by $q_s$) performed by the adversary. We work in the random oracle model, and
assume hardness of the Decisional Diffie Helman problem.

We use the following notation throughout the paper. In the description of the
protocols, $ac$ is the authentic channel, $pc$ is the private channel, and $bc$
is the broadcast channel. Assignment is denoted by $\assign$, and
$\overset{R}{\leftarrow}$ means selecting an element uniformly at random from
the indicated set. Receiving messages from the channel or the broadcast network
can be done in a blocking fashion (indicated by \keyw{receive}) or
in a non-blocking fashion (indicated by \keyw{on\ receiving}).

In message flowcharts, \sendright{m} denotes sending $m$ on the private or 
authentic channel, while  \broadcastright{m} denotes broadcasting $m$
on the broadcast channel. The receiving party puts the message in the indicated
variable $v$ at the arrowhead. 

%$\encrypt{}{m}$: the message $m$ encrypted with $k$ using a symmetric cipher.

\section{Ephemeral key exchange protocols}
\label{sec-exchange}

In this section we present \eke{} protocols, for varying assumptions on the
properties of the low bandwidth channel between Alice and Bob. We start with
the case where the channel between Alice and Bob is unidirectional and
private as well as authentic. Then we discuss the case where the channel is
bidirectional. We present a protocol for just private channels, and finish
with a protocol where the channel is only authentic.

In some of the protocols, an EKE protocol~\cite{BelM92,KatOY01} is used as the
basic building block. This EKE protocol is assumed to broadcast its messages
over the broadcast channel instead of sending them point to point.

\subsection{\eke{} for an unidirectional private and authentic channel}

In the unidirectional private and authentic channel case, existing EKE
protocols can easily be used as a building block. The channel is simply
used to reliably send a random password from the client to the server, after
which the EKE protocol is run to exchange the key. This
is laid down in Prot.~\ref{prot-privauth}. 
The security parameters are set by $\sps=\cap$ and $\spl=\sks$.

\begin{protocol}
\hspace{\protindent}
\begin{minipage}{\textwidth}
\begin{tabbing}
\IF\= client \\
\> \THEN\= $p \overset{R}{\leftarrow} \Set{0}{2^\sps-1}$ \\
\> \>   \keyw{send} $p$ \keyw{on} $pc$ \\
\> \ELSE \keyw{receive} $p$ \keyw{from} $pc$ \\
$k \assign \text{EKE}(p)$ 
\end{tabbing}
\end{minipage}
\caption{\eke{} for unidirectional private and authentic channel.}
\label{prot-privauth}
\end{protocol}

\subsubsection{Analysis}

We assume the underlying EKE protocol is correct and secure.
If Alice and Bob want to exchange a key, it is straightforward to show that in
an honest execution of Prot.~\ref{prot-privauth}, at the end of the protocol
they do actually share the same key.

Next we show this protocol is secure.  
\begin{theorem}
\label{th-privauth}
The advantage of an adversary attacking Prot.~\ref{prot-privauth} is at most
the advantage of any adversary attacking the basic EKE protocol.
\end{theorem}
\begin{proof}
Suppose an adversary attacks a run of protocol Prot.~\ref{prot-privauth} with
advantage $a$. Because by assumption, the adversary cannot control or gain
information from the messages sent over the private and authentic channel, the
advantage of the adversary would still be $a$ when given this run where all
messages sent over the channel are random, independent, values. 
But this is a run over the basic EKE protocol, with additional random values
added to it. Hence the adversary can attack the basic EKE protocol with
advantage $a$ by adding random values to it and treating it as a run over
Prot.~\ref{prot-privauth}.
\qed
\end{proof}
Note that each execution of the EKE protocol is given a fresh password. This is
unlike the typical case for EKE protocols, where each pair of nodes use the
same password each time they wish to connect. This negatively impacts the upper
bound for \eke{} protocols on the advantage of the adversary, in that the
advantage of the adversary increases too quickly with the number of times he
tries to guess the password. Because Prot.~\ref{prot-privauth} uses a fresh
password for each execution of the EKE protocol, the upper bound could be
improved slightly if we consider one particular instance of an EKE protocol in
our analysis.

\subsection{\eke{} for a bidirectional private channel}

If the channel is bidirectional and private (without being authentic),
existing EKE protocols can also be used as a building
block.  If the channel is bidirectional, Alice and Bob
simply generate two short $\sps$ bit passwords, exchange them over the private
channel, and subsequently run an EKE protocol
using the exclusive OR\footnote{
  Using the exclusive OR instead of concatenation makes the resulting EKE
  password as long as the \eke{} short security parameter. Moreover, it makes
  the protocol for Alice and Bob symmetric.
}
of both passwords as the EKE password to establish the
shared session key. Security of this protocol is based on the observation that
although anybody can try to set up a session with Bob by sending him a
password, Bob will only divulge his own password to the person he wants to
connect to, \ie{} Alice. Therefore, only Alice is capable of generating the EKE
password that will be accepted by Bob. In other words, Alice's authenticity is
verified by the fact that she knows Bob's password. 
The protocol is detailed in Prot.~\ref{prot-biprivate}. 
Again, the security parameters are set by $\sps=\cap$ and $\spl=\sks$.

\begin{protocol}
\hspace{\protindent}
\begin{minipage}{\textwidth}
\begin{tabbing}
+++\=+++\=+++\=\kill
$p \overset{R}{\leftarrow} \Set{0}{2^\sps-1}$ \\
\keyw{send} $p$ \keyw{on} $pc$ \\
\keyw{receive} $q$ \keyw{from} $pc$ \\
$r \assign p \xor q$ \\
$k \assign \text{EKE}(r)$
\end{tabbing}
\end{minipage}
\caption{\eke{} for bidirectional private channel.}
\label{prot-biprivate}
\end{protocol}

\subsubsection{Analysis}

It is again straightforward to show that if Alice and Bob want to exchange a 
key using Prot.~\ref{prot-biprivate}, they will actually share the same key in
an honest execution thereof, if we assume the underlying EKE protocol is
correct.

Next we prove security of the protocol.
\begin{theorem}
The advantage of an adversary attacking Prot.~\ref{prot-biprivate} is at most
the advantage of any adversary attacking the basic EKE protocol.
\end{theorem}
\begin{proof}
Suppose in a run of Prot.~\ref{prot-biprivate}, an adversary attacks this run
with advantage $a$. The password used by an instance depends on a value
received on the private channel, xor-ed with a private random value that is
also sent privately to the other party. Because by assumption the adversary
cannot gain information from the messages sent over the private channel,
the password used in an instance of the basic EKE protocol is independent of
the values exchanged over the private channel. Hence by similar reasoning as in
theorem~\ref{th-privauth}, the advantage of the adversary attacking the basic
EKE protocol is at least $a$.
\qed
\end{proof}

\subsection{\eke{} for a bidirectional authentic channel}

For the \eke{} protocol for a bidirectional authentic channel we use a
different approach, not using an EKE protocol as the basic building block.
The idea behind the protocol (presented as protocol~\ref{prot-biauth})
is the following. 

\begin{protocol}[t]
\hspace{\protindent}
\begin{minipage}{\textwidth}
\begin{tabbing}
+++\=+++\=+++\=\kill
\textit{Commit} \\

\> pick random $x$ \\
\> \keyw{broadcast} $h_1(\modp{g^x})$ \keyw{on} $bc$ \\
\> \keyw{receive} $\alpha$ \keyw{from} $bc$ \\

\textit{Authenticate} \\

\> \keyw{send} $h_2(\modp{g^x})$ \keyw{on} $ac$ \\
\> \keyw{receive} $\beta$ \keyw{from} $ac$ \\

\textit{Key exchange} \\

\> \keyw{broadcast} $\modp{g^x}$ \keyw{on} $bc$ \\
\> \keyw{receive} $m$ \keyw{from} $bc$ \\
\> \IF $h_1(m) = \alpha$ and $h_2(m) = \beta$ \\
\> \> \THEN $u \assign m$\\
\> \> \ELSE \keyw{abort} \\

\textit{Key validation} \\

\> $j \assign \begin{cases}
        0 & \text{if client} \\
        1 & \text{if server}
        \end{cases}$ \\
\> \keyw{broadcast} $h_{4+j}(\modp{u^x})$ \keyw{on} $bc$ \\
\> \keyw{receive} $m$ \keyw{from} $bc$ \\
\> \IF $h_{5-j}(\modp{u^x}) = m$ \\
\> \> \THEN $k = h_3(\modp{u^x})$ \\
\> \> \ELSE \keyw{abort} \\
\end{tabbing}
\end{minipage}
\caption{\eke{} for bidirectional authentic channel.}
\label{prot-biauth}
\end{protocol}

\begin{figure}[t]
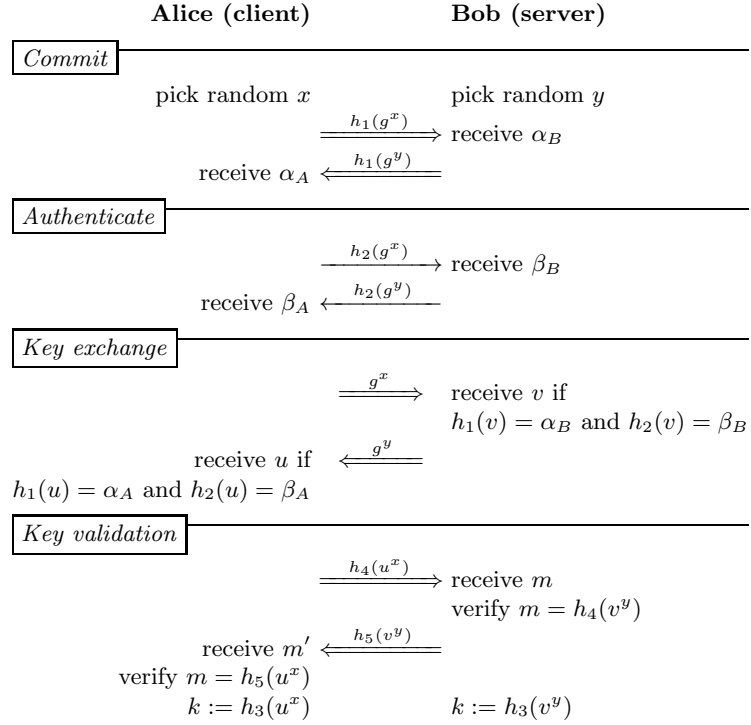

\small
\begin{center}
\begin{tabular}{rcl}
\textbf{Alice (client)} & & \textbf{Bob (server)} \\

\phase{Commit}

pick random $x$    &                      & pick random $y$ \\
                   & \broadcastright{h_1(\modp{g^x})} 
                                          & receive $\alpha_B$ \\
receive $\alpha_A$        & \broadcastleft{h_1(\modp{g^y})}  
                                          & \\ 

\phase{Authenticate}

                   & \sendright{h_2(\modp{g^x})}
                                          & receive $\beta_B$ \\
receive $\beta_A$   & \sendleft{h_2(\modp{g^y})}
                                          & \\
\phase{Key exchange}

                   & \broadcastright{\modp{g^x}}
                                          & receive $v$ if \\
                   &                      & $h_1(v) = \alpha_B$  
                                            and $h_2(v) = \beta_B$ \\ 
receive $u$ if  & \broadcastleft{\modp{g^y}}
                                          & \\
$h_1(u) = \alpha_A$  
and $h_2(u) = \beta_A$
                   &                      & \\

\phase{Key validation}

                & \broadcastright{h_4(\modp{u^x})}
					& receive $m$ \\
		&			& verify $m = h_4(\modp{v^y})$ \\
receive $m'$	& \broadcastleft{h_5(\modp{v^y})} \\
verify $m=h_5(\modp{u^x})$ \\
$k  \assign h_3(\modp{u^x})$
                &                       & $k \assign h_3(\modp{v^y})$ 
\end{tabular}
\end{center}
\caption{Message flow of \eke{} for a bidirectional authentic channel.}
\label{fig-biauth}
\end{figure}

To establish a shared session key, Alice and Bob will use a Diffie-Helman type
key exchange~\cite{DifH76}. To avoid man-in-the-middle attacks, the
shares must be authenticated. However, the capacity of the
authentic channel is too small to do so directly. Instead, Alice and Bob
proceed in four phases. In the first phase (the \term{commit phase}) Alice and
Bob commit to their shares without revealing them. Then in the
\term{authentication phase} they will send a small authenticator of their share
to each other 
over the authentic channel.  In the \term{key exchange phase}, both will reveal
their share.  Only shares committed to will be accepted, and the share matching
the authenticator will be used to compute the shared session key. The key is
verified in the final \term{key validation phase} to ensure that Alice and
Bob indeed share the same session key, using the mechanism described
in~\cite{BelPR00}. Only if the validation phase is successful the protocol will
accept.

Note that we must first commit to a value before revealing either the value or
the authenticator, or else the adversary can trivially (in an expected
$2^{\cap-1}$ number of tries) find a share of his own that matches the
authenticator that will be sent by Alice. 

In Prot.~\ref{prot-biauth}, the security parameters are determined by the 
size of the session key established and the capacity of the authentic channel.
We set $\spl=\sks$ and $\sps=\cap$.
$G$ is a group of order at least $2^{2\spl}$ 
with generator $g$ for which the Decisional Diffie Helman (DDH) problem is
hard. A possible candidate is the subgroup of order $q$ in $\Z_p^*$ for $p,q$
prime and $p=2q+1$~\cite{Bon98}.
Naturally, exponentiations like $g^x$ are computed in the group $G$.

Furthermore, we use two hash functions
$h_1 : G \mapsto G$ and $h_2 : G \mapsto \Set*{0,1}^\cap$, that
satisfy the following property.
\begin{property}
\label{prop-indep}
Let $X$ be a uniformly distributed random variable over $G$, and let 
$a \in \Set*{0,1}^\cap$ and $b \in  G$ be arbitrary. We assume that the
two hash functions $h_1,h_2$ satisfy
\[
   \CProb{h_2(X) = a}{h_1(X) = b} = \Prob{h_2(X)=a} = 2^{-\cap}~.
\]
\end{property}
Finally, pairwise independent hash functions 
$h_3,h_4,h_5 : G \mapsto \Set*{0,1}^\sks$ are used as well.
In practice, these hash functions can be derived from a single hash function
$h$ using the equation $h_i(x) = h(x \concat i)$ (where $\concat$ denotes
concatenation of bit strings).

\subsubsection{Analysis}

It is straightforward to show that in an honest execution of
Prot.~\ref{prot-biauth}, 
if Alice and Bob want to exchange a key, at the end of the protocol they do
actually share the same key.

Security of Prot.~\ref{prot-biauth} is proven as follows.
We use the following result presented by Boneh~\cite{Bon98}, which holds under
the assumption that the Decisional Diffie Helman problem over $G$ is hard.
\begin{proposition}
\label{prop-ddh}
Let the order of $G$ be at least
$2^{2\spl}$, and let $h_3 : G \mapsto \Set*{0,1}^\spl$ be a
pairwise independent hash function. Then the advantage of any adversary
distinguishing $h_3(g^{ab})$ from a random element of $\Set*{0,1}^\spl$, when
given $g^a,g^b$ is a most $O(2^{-\spl})$.
\end{proposition}
Using this proposition we are able to prove the following theorem.
\begin{theorem}
The advantage of an adversary attacking Prot.~\ref{prot-biauth} using at most
$q_{\text{send}}$ send queries is at most 
\[
O(1-e^{-q_{\text{send}}/2^{\sps}}) + O(2^{-\spl})~.
\]
\end{theorem}
\begin{proof}
We split the proof in two cases. We first consider the case where the session
key $k$ generated by an oracle is not based on a share $\modp{g^a}$
sent by the adversary and derived from a value $a$ of his own choosing, and
then consider the case where the adversary manages to convince the oracle to
use such a share of his own choosing.

If the session key generated by an oracle is not based on a share $\modp{g^a}$
sent by the adversary and derived from a value $a$ of his own choosing, then
$k$ depends on private random values $x,y$ unobserved by the adversary and
publicly exchanged shares $g^x$ and $g^y$ using a Diffie-Helman (DH) key
exchange. Any adversary attacking Prot.~\ref{prot-biauth} can be converted to
an adversary attacking a basic DH key exchange, by inserting the necessary
hashes $h_i(g^x)$ and $h_j(g^y)$ (for $i,j \in \Set*{1,2,3}$) and random values
for $h_4()$ and $h_5()$ (this is possible due to the random oracle model and
Prop.~\ref{prop-ddh}) in the run of the basic DH key exchange before analysing
the run. Hence the advantage of the adversary to distinguish the session key
cannot be higher than its advantage in breaking the Diffie-Helman key exchange,
which is at most $O(2^{-\spl})$ by Prop.~\ref{prop-ddh}.

In the other case, in order
to convince an oracle of $A$ to use the share $\modp{g^a}$ of the adversary
in the third phase of the protocol, the adversary must ensure that
\begin{itemize}
\item $h_1(\modp{g^a}) = \alpha$, and
\item $h_2(\modp{g^a}) = \beta$
\end{itemize}
for values $\alpha, \beta$ used in this oracle. Note that $\beta$ is unknown in
the commit phase. Moreover, property~\ref{prop-indep} guarantees it is
independent of values exchanged during the commit phase. Therefore, for any
value $\modp{g^a}$ committed by the adversary in the commit phase, the
probability that $h_2(\modp{g^a}) = \beta$ is $2^{-\cap}$. 

For each send query then the probability of success is $2^{-\cap}$. Success
with one instance is independent of success in any other instance. Hence, with
$q_{\text{send}}$ send queries, the probability of success becomes 
(cf.~\cite{Fel57})
\[ 
1-(1-2^{-\cap})^{q_{\text{send}}} \approx 1-e^{-2^{-\cap}q_{\text{send}}}
\]
With $t=\cap$ this proves the theorem.
\qed
\end{proof}
Note that in fact the advantage of the adversary attacking the \eke{} protocol
is strictly less than the advantage of the adversary attacking password based
EKE protocols, like the protocol of Katz \etal~\cite{KatOY01} whose
advantage is bounded by 
\[ 
   O(q_{\text{send}} / 2^\sps) + O(2^{-\spl})~,
\]
where, loosely speaking, $q_{\text{send}}$ is the number of times the adversary
tries to guess the password. The difference is caused by the fact that in the
EKE setting, multiple instances of the protocol use the same 
password\footnote{
  This could be overcome by allowing only the first $z$ connections to use the
  password alone, and using parts of the previously established shared secrets
  to generate new, longer, passwords. Then the bound on the advantage of the
  adversary essentially becomes equal to ours.
}.

\section{Applications}
\label{sec-appl}

Beyond those mentioned in the introduction, there are many other situations
that involve ephemeral pairing. 
\begin{itemize}
\item Connecting two  laptops over an infrared connection, while in a business
  meeting.  
\item Buying tickets wirelessly at a box office, or verifying them at the
  entrance. 
\item Unlocking doors using a wireless token, making sure the right door is
  unlocked. 
\end{itemize}
For all these applications it is very important that the burden of correctly
establishing the right pairing should not solely rest on the user. The user may
make mistakes, and frequent wrong pairings will decrease the trust in the
system. This is especially important for applications that involve financial
transactions. On the other hand, some user intervention will obviously always
be required. The trick is to make the user actions easy and intuitive given the
context of the pairing.

In the next section, we describe how a low bandwidth authentic or
private channel can be implemented in quite practical settings. These are of
course merely suggestions. There are probably many more and much better ways to
achieve the same effect. The point here is, however, to merely show that such
channels can be built in principle.

\subsection{Implementing the low bandwidth authentic or private channel}

To implement an authentic or private channel in practice,
several solution stra\-te\-gies are applicable. 
\begin{itemize}
\item Establishing physical contact, either by a wire, through a connector, or
  using proximity techniques.
\item Using physical properties of the wireless communication link, that may
  allow `aiming' your device to the one you wish to connect to.
\item Using fixed visible identities, either using explicitly shown unique
  names on devices, or using the unique appearance of each device.
\item Using small displays that can either be read by the operator of the other
  device or read directly by the other device.
\end{itemize}
Which strategy to select depends very much on the specific application
requiring ephemeral pairing.
We will discuss each of these strategies briefly.

\subsubsection{Physical contact}

The easiest way to solve ephemeral pairing is to connect both nodes
(temporarily) physically, either by a wire, or by making them touch each others
conductive pad. The resulting physical connection can be used as the private or
authentic channel in the previous protocol. Or it can be used to
exchange the shared secret directly, of course

\subsubsection{Fixed visible identities}

Here one could use for example numbers, or the physical appearance. Each node
holds a unique private key, and the physical identity is bound to the
corresponding public key using a certificate generated by the certification
authority (CA) managing the application.

The main drawback is that these solutions require a central Certification
Authority. Moreover, the a priori distribution of secrets is contrary to the
spirit of the \eke{} problem.

A variant (described in~\cite{MeT01}) uses the fixed identity of nodes in the
following way. Any node wishing to connect can do so. Each connection is
assigned a unique and small connection number, which is shown on a display. 
The user mentions the number to the merchant, who then initiates a payment
over the indicated connection. The problem with this setup is that it is
vulnerable to man-in-the-middle attacks.

\subsubsection{Physical link properties}

Depending on the properties of the physical link, one could reliable aim a
device at another, or safely rule out connections to/from other far away
devices. 

\subsubsection{Operator read displays}

In this scheme, each node has a small display and a way to select several
images or strings from the display (through function keys or using a
touch pad). An authentic channel can be implemented as follows. To send a $\cap$
bit string, it is converted to a simple pattern that is shown on the display
of the sender. The receiver enters the pattern on its device, which converts it
back to the $\cap$ bits.

\section{Conclusions}
\label{sec-concl}

We have formulated the ephemeral pairing problem, and have presented several
ephemeral key exchange protocols showing that this problem can be solved using
small capacity, and mostly bidirectional, point to point channels and a
broadcast network with identities to separate communication streams.

More work needs to be done to develop \eke{} protocols using only
unidirectional channels, or on truly anonymous broadcast networks.

It would be interesting to develop protocols that are correct under less 
strong assumptions, \ie ones that do not require to assume either the 
random oracle model or hardness of the Decisional Diffie Helman problem (or
both). The same holds for the assumption on the authentic channel that
adversary cannot modify or inject messages of his choice at all. More research
is needed to investigate the effects on the advantage of the adversary
if he can modify or inject messages on the authentic channel with 
low success probability. 

\section{Acknowledgements}

This work was inspired by the work of Yan Yijun on a payments architecture for
mobile systems, under the supervision of Leonard Franken of the ABN AMRO bank
and myself. I would like to thank Yan and Leonard for fruitful initial
discussions on this topic.

\bibliography{pairing}

\begin{thebibliography}{{Mob}01}

\bibitem[BPR00]{BelPR00}
{\sc Bellare, M., Pointcheval, D., and Rogaway, P.}
\newblock Authenticated key exchange secure against dictionary attacks.
\newblock In {\em \bibselect{EUROCRYPT} {Advances in Cryptology --- EUROCRYPT}
  2000\/} (Bruges, Belgium, 2000), B.~Preneel (Ed.), \bibselect{LNCS} {Lect.\
  Not.\ Comp.\ Sci.\ } 1807, Springer, pp.~139--155.

\bibitem[BM92]{BelM92}
{\sc Bellovin, S.~M., and Merritt, M.}
\newblock Encrypted key exchange: Password-based protocols secure against
  dictionary attacks.
\newblock In {\em \bibselect{IEEE Security \& Privacy} {IEEE Symp.\ on Security
  and Privacy}\/} (Oakland, CA, USA, 1992), IEEE, pp.~72--84.

\bibitem[BM93]{BelM93}
{\sc Bellovin, S.~M., and Merritt, M.}
\newblock Augmented encrypted key exchange: A password-based protocol secure
  against dictionary attacks and password file compromise.
\newblock In {\em 1st \bibselect{CCS} {Int.\ Conf.\ on Computer and
  Communications Security}\/} (Fairfax, VA, USA, 1993), ACM, pp.~244--250.

\bibitem[Bon98]{Bon98}
{\sc Boneh, D.}
\newblock The decision {Diffie-Hellman} problem.
\newblock In {\em Proc.\ of the 3rd Algorithmic Number Theory Symp.\/} (1998),
  \bibselect{LNCS} {Lect.\ Not.\ Comp.\ Sci.\ } 1423, pp.~48--63.

\bibitem[BMP00]{BoyMP00}
{\sc Boyko, V., MacKenzie, P., and Patel, S.}
\newblock Provably secure password-authenticated key exchange using
  {Diffie-Hellman}.
\newblock In {\em \bibselect{EUROCRYPT} {Advances in Cryptology --- EUROCRYPT}
  2000\/} (Bruges, Belgium, 2000), B.~Preneel (Ed.), \bibselect{LNCS} {Lect.\
  Not.\ Comp.\ Sci.\ } 1807, Springer, pp.~156--171.

\bibitem[CK01]{CanK01}
{\sc Canetti, R., and Krawczyk, H.}
\newblock Analysis of key-exchange protocols and their use for building secure
  channels.
\newblock In {\em \bibselect{EUROCRYPT} {Advances in Cryptology --- EUROCRYPT}
  2001\/} (Innsbruck, Austria, 2001), B.~Pfitzmann (Ed.), \bibselect{LNCS}
  {Lect.\ Not.\ Comp.\ Sci.\ } 2045, Springer, pp.~453--474.

\bibitem[{C}ep01]{CEPS01}
{\sc {C}eps{C}o}.
\newblock {C}ommon {E}lectronic {P}urse {S}pecifications: Technical
  specification, version 2.3, 2001.
\newblock \texttt{http://www.cepsco.com}.

\bibitem[DH76]{DifH76}
{\sc Diffie, W., and Hellman, M.~E.}
\newblock New directions in cryptography.
\newblock {\em IEEE Trans. Inf. Theory {\bf IT-11}\/} (1976), 644--654.

\bibitem[Fel57]{Fel57}
{\sc Feller, W.}
\newblock {\em An Introduction to Probability Theory and Its Applications},
  2nd~ed.
\newblock Wiley \& Sons, New York, 1957.

\bibitem[GL03]{GenL03}
{\sc Gennaro, R., and Lindell, Y.}
\newblock A framework for password-based authenticated key exchange.
\newblock Tech. rep., IBM T.J. Watson, 2003.
\newblock Abstract appeared in EUROCRYPT 2003.

\bibitem[Jab96]{Jab96}
{\sc Jablon, D.~P.}
\newblock Strong password-only authenticated key exchange.
\newblock {\em \bibselect{Comput.\ Comm.\ Rev.} {Computer Communications
  Review}\/} (1996).
\newblock \texttt{http://www.std.com/\char`~dpj} and
  \texttt{www.integritysciences.com}.

\bibitem[JW01]{JurW01}
{\sc Jürjens, J., and Wimmel, G.}
\newblock Security modelling for electronic commerce: The common electronic
  purse specifications.
\newblock In {\em First IFIP conference on e-commerce, e-business, and
  e-government (I3E)\/} (Zürich, Switzerland, 2001), Kluwer.

\bibitem[KOY01]{KatOY01}
{\sc Katz, J., Ostrovsky, R., and Yung, M.}
\newblock Efficient password-authenticated key exchange using human-memorable
  passwords.
\newblock In {\em \bibselect{EUROCRYPT} {Advances in Cryptology --- EUROCRYPT}
  2001\/} (Innsbruck, Austria, 2001), B.~Pfitzmann (Ed.), \bibselect{LNCS}
  {Lect.\ Not.\ Comp.\ Sci.\ } 2045, Springer, pp.~475--494.

\bibitem[Luc97]{Luc97}
{\sc Lucks, S.}
\newblock Open key exchange: How to defeat dictionary attacks without
  encrypting public keys.
\newblock In {\em The Security Protocol Workshop '97\/} (1997), pp.~79--90.

\bibitem[Mac01]{Mac01a}
{\sc MacKenzie, P.}
\newblock More efficient password-authenticated key exchange.
\newblock In {\em Topics in Cryptography (CT-RSA)\/} (2001), \bibselect{LNCS}
  {Lect.\ Not.\ Comp.\ Sci.\ } 2020, Springer, pp.~361--377.

\bibitem[{Mob}01]{MeT01}
{\sc {Mobile electronic {T}ransactions}}.
\newblock Solution to {B}luetooth multiuser problem using method of tokens.
\newblock Tech. rep., 2001.
\newblock \texttt{http://www.mobiletransaction.org/}.

\bibitem[Sho99]{Sho99}
{\sc Shoup, V.}
\newblock On formal models for secure key exchange.
\newblock Tech. Rep. RZ 3120 (\#93166), IBM, 1999.
\newblock Invited talk at ACM Computer and Communications Security conference,
  1999.

\bibitem[SA99]{StaA99}
{\sc Stajano, F., and Anderson, R.}
\newblock The resurrecting duckling: Security issues for ad-hoc wireless
  networks.
\newblock In {\em Security Procotols, 7th Int. Workshop\/} (1999),
  B.~Christianson, B.~Crispo, and M.~Roe (Eds.), \bibselect{LNCS} {Lect.\ Not.\
  Comp.\ Sci.\ }, pp.~172--194.

\end{thebibliography}

\end{document}